\documentclass[12pt]{article}
\usepackage[T1]{fontenc}
\usepackage{tgpagella}
\usepackage{amsmath,amssymb,mathtools}
\usepackage{graphicx,epsf,color,xcolor}
\usepackage{enumerate}
\usepackage{natbib}
\usepackage{forest}
\usepackage{float}
\usepackage{url} 
\usepackage{times}
\usepackage[ruled,noend]{algorithm2e}
\usepackage{graphics}
\usepackage{booktabs}
\usepackage{xr}
\usepackage[colorlinks=true, linkcolor=blue, citecolor=blue, filecolor=blue]{hyperref}
\usepackage{setspace}
\usepackage{caption}
\usepackage{subcaption}

\newcommand{\blind}{0}
\usepackage[margin=1in]{geometry}

\usepackage{times}
\usepackage{lineno,hyperref}

\usepackage{natbib}
\usepackage{caption}
\usepackage{algpseudocode}
\usepackage{soul}
\newtheorem{theorem}{Theorem}[section]
\newtheorem{corollary}{Corollary}[theorem]

\usepackage{bm}
\newcommand{\mT}{\mathcal{T}}
\modulolinenumbers[5]
\usepackage{enumitem}

\newif\ifanon
\anonfalse

\begin{document}

\def\spacingset#1{\renewcommand{\baselinestretch}%
{#1}\small\normalsize}
\spacingset{1}

\title{\bf Identifying World Events in Dynamic International Relations Data Using a Latent Space Model}

\if1\blind
{
  \author{}\date{}
  \maketitle
} \fi

\if0\blind
{
\author{Yunran Chen and Alexander Volfovsky\thanks{This research was partially supported by grants from the National Science Foundation (CAREER DMS-2046880); the Army Research Institute (W911NF-18-1-0233). Corresponding author; alexander.volfovsky@duke.edu}\\
\textit{Department of Statistical Science, Duke University, USA}\\}
\maketitle
}
\fi

\begin{abstract}
Dynamic network data have become ubiquitous in social network analysis, with new information becoming available that captures when friendships form, when corporate transactions happen and when countries interact with each other. Flexible and interpretable models are needed in order to properly capture the behavior of individuals in such networks. In this paper, we focus on study the underlying latent space that describes the social properties of a dynamic and directed international relations network of countries. We extend the directed additive and multiplicative effects network model to the continuous time setting by treating the time-evolution of model parameters using Gaussian processes. Importantly we incorporate both time-varying covariates and node-level additive random effects that aid in increasing model realism. We demonstrate the usefulness and flexibility of this model on a longitudinal dataset of formal state visits between the world's 18 largest economies. Not only does the model offer high quality predictive accuracy, but the latent parameters naturally map onto world events that are not directly measured in the data.
\end{abstract}


\newpage

\section{Introduction}

Encoding relationship information between individuals, firms or countries, can be invaluable to understanding disease transmission \citep{bu2020likelihood}, financial transactions \citep{durante2014bayesian}, and international relations \citep{hoff2004modeling}. The actors in these relationships can be represented as nodes or vertices in a graph and the interactions between them as edges. While early data on networks was static \citep{sampson1968novitiate,zachary1977information,harris2019cohort}, modern health, international relations and social science applications feature evolving, or dynamic, network information. These rich new data require the extension of previously static models and computational tools to include dynamics\citep{butts_lomi_snijders_stadtfeld_2023}. In this work we extend a dynamic latent space model for directed networks that generalizes the class of latent space models for static directed networks. 

Static latent space network models have been successful at capturing important social processes \citep{hoff2005bilinear,hoff2008modeling,hoff2009multiplicative}: (i) popularity -- a node is more likely to be nominated as a connection than other nodes; (ii) sociability -- a node is more likely to nominate a large number of connections; (iii) homophily and stochastic equivalence -- nodes that are similar on some observed or unobserved attributes have similar network connections; and (iv) reciprocity -- the presence of a directed edge between two individuals means there is more likely to be a reciprocal directed edge between them. It is natural to develop dynamic models that can also capture these, and possibly to allow for these concepts to adapt or change over time. 

\begin{figure}
\includegraphics[width=\linewidth]{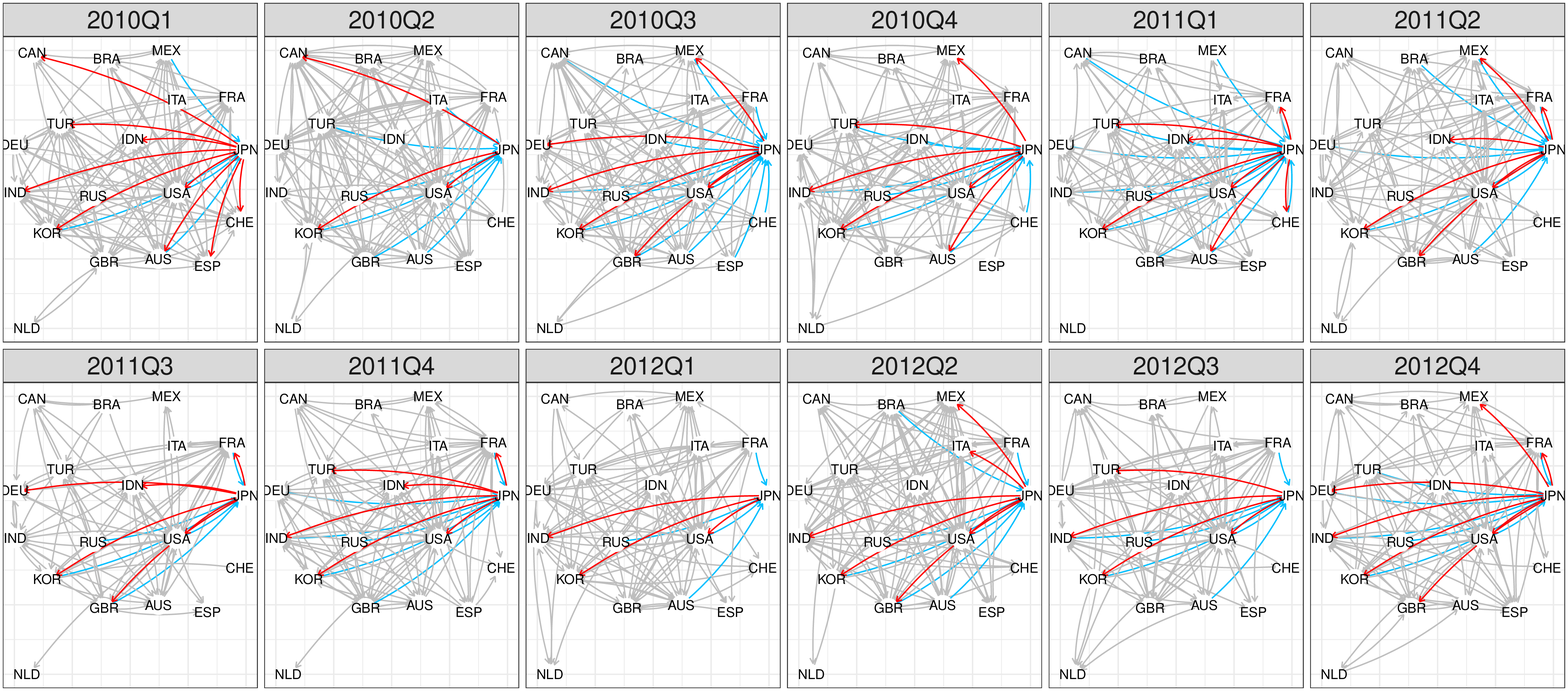}
\caption{Observed networks for selected time windows. Each edge represents at least one visit within a quarter. We highlight Japan's visiting activities as a visitor and a host using red and blue color respectively. Temporal reciprocity, nodal heterogeneity and dynamic evolution of the directed network can be found in the dynamic networks.}
\label{fig:net}
\end{figure}

To motivate this desiderata we plot the official visits across the world's top 18 largest economies, excluding China and Saudi Arabia, from 2010 to 2012 in Fig.~\ref{fig:net}\footnote{Data are from the Integrated Crisis Early Warning System(ICEWS) dataverse and described in detail in Section~\ref{sec:app}}. Each directed edge represents at least one official visit from one country to another within a specified quarter. 
 
We can observe all four aforementioned social properties: Russia visits many other countries (suggesting high sociability) but is not visited by many (suggesting low popularity). The United States and the United Kingdom share many global goals and hence exhibit similar behavior (suggesting homophily and stochastic equivalence). Reciprocity is a bit more complex: for example there are many simultaneous edges between the United States and Japan, but it is likely that visits trigger reciprocity in future time points. As such, we argue that temporal reciprocity is a more meaningful concept: a visit by Japan in Q1 2010 to Turkey, the United Kingdom and Australia appears to be reciprocated by these countries to Japan in Q2 2010. We will return to discussing Japan in our detailed data analysis in Section \ref{sec:app}.

Beyond explaining observed behaviors, statistical models are frequently tasked with establishing connections between collected data and unobserved information. It is evident that international relations are driven by a myriad of events, including natural disasters, in-country political positions, financial crises, among others. We will demonstrate that the dynamically changing latent variables in our model are correlated with, that is predictive of or can predict, real-world events that are not included in the data that the model is based. This provides an invaluable resource for researchers attempting to answer ``what if?'' questions as some interactions between network events and world events may remain unnoticed without a detailed analysis of the network. This can be illustrated by interactions between Japan and France (that we study in detail in Section~\ref{sec:app}): After the Fukushima disaster, the probability of official visits to Japan decreased for all countries, except for France and the USA, where the probability of visits increased. This unpredicted behavior both demonstrates France's long-term support for Japan and hints at specific interactions between France and Japan on nuclear development (even though the plant used American made reactors). This highlights the significance of dynamic network analysis in uncovering unexpected trends and relationships.

The most direct approach to dynamic network modeling is to consider snapshots at different time points and model those separately, potentially ignoring the temporal information \citep{ward2007persistent,ward2007disputes}. By assuming a first order Markovian property, which indicates networks only depend on the information at prior time points, these models can be extended to include generalized autoregressive terms such as a lag one response variable or a lag one latent factor \citep{ward2013gravity,minhas2016new,hoff2015multilinear,he2019multiplicative}. Another commonly used models for dynamic networks are the temporal exponential random graph model (TERGM) \citep{hanneke2010discrete,krivitsky2014separable,lee2020model} and the stochastic actor-oriented model (SAOM) \citep{snijders1996stochastic,snijders2001statistical,snijders2005models,snijders2010introduction,uzaheta_amati_stadtfeld_2023} which largely model changes to network summary statistics or require an actor-driving mechanism. Other work incorporates a transition matrix or a state space model of latent factors or nodal characteristics into static network models \citep{xing2010state,ho2011evolving,yang2011detecting,xu2013dynamic,sewell2015latent}.
These models require imposing strong assumptions about how information is shared between nodes and across time-scales, but modern data streams require greater flexibility. 

For undirected networks, the greatest modeling flexibility is provided by 
\cite{durante2014nonparametric} and \cite{kim2018dynamic}. Both propose versions of Gaussian Process priors for a latent space representation of the network. In this article we describe the extension of this modeling approach to directed networks that are additionally able to account for nodal heterogeneity. Importantly, the only prior dependence that the models for undirected networks accounted for was temporal dependence, but in directed networks the actions of individuals as senders and receivers of connections are likely correlated, which is also incorporated into our priors. We provide a short primer for dynamic latent space models in Section~\ref{sec:modelall} (that adapts a Gibbs sampler with P\`olya-Gamma augmentation \citep{polson2013bayesian} to directed networks) and then turn our attention in Section~\ref{sec:app} to the analysis of the official visits networks of Fig.~\ref{fig:net} with an in-depth case study of Japan. We identify how various model parameters capture meaningful and relevant behaviors in international relations and map those onto world events (such as economic downturns and natural disasters) that are not directly measured in the data. We relegate model validation, sensitivity to misspecification and sensitivity to prior choice simulation studies to Section \ref{sec:sim}. We conclude with a discussion of future direction of both methods and international relations data in Section~\ref{sec:discussion}.

\section{Dynamic Latent Space Model}
\label{sec:modelall}
Our model extends the directed additive and multiplicative model of \cite{hoff2015dyadic} to the dynamic case. Throughout we consider a network between $n$ units or nodes. We let $y_{ij}(t)$ be a binary indicator of an interaction between unit $i$ and unit $j$ at time point $t\in\mT$ and $\pi_{ij}(t)=\Pr(y_{ij}(t)=1)$ be the probability of that interaction. The collection of interactions at time point $t$ is frequently referred to as the adjacency matrix at time $t$ and is denoted by $Y(t)=\{y_{ij}(t),i\neq j\}$ where the diagonal entries are not defined by convention. Note that since we are considering asymmetric relationships, it is not necessarily the case that $y_{ij}(t)=y_{ji}(t)$ (or that $\pi_{ij}(t)=\pi_{ji}(t)$). In addition to observing the interactions among units we can observe temporally changing unit and dyad specific covariates (in the dataset we discuss this covariate is the GDP of each country) that are collectively denoted by $Z_{ij,t}$. We can define the full dynamic network model as follows:
\begin{align}
    y_{ij}(t)|\pi_{ij}(t) &\sim Bernoulli\{\pi_{ij}(t)\}
    \label{equ:ber}    \\
    \pi_{ij}(t) &=\{1+e^{-S_{ij}(t)}\}^{-1}
    \label{equ:logistic}\\
    S_{ij}(t) &=\mu(t)+Z_{ij,t}^T\beta(t) + {x_i^s}^T(t)x^r_j(t) + a_i(t) + b_j(t).
    \label{S_equ}
\end{align}

This proposed dynamic model is a generalized version of the undirected models proposed by \citet{durante2014bayesian,durante2014nonparametric}.

In our model $\beta(t)$ is a vector of time-varying coefficients associated with the observed covariates. The remaining elements in Eq.~\eqref{S_equ} are latent additive and multiplicative effects \citep{hoff2015dyadic}: The dynamic additive effects $a_i(t)$ and $b_j(t)$ capture the ``sociability'' of unit $i$ at time $t$ and ``popularity'' of unit $j$ at time $t$ respectively, and the $H$-dimensional multiplicative effects $x_i^s(t)$ and $x_j^r(t)$ capture temporal third order dependence in directed relationships from unit $i$ to unit $j$ at time $t$. In general these higher order effects capture notions of homophily and stochastic equivalence which are frequently observed in network data, while their temporal structure allows for the levels of homophily and stochastic equivalence among units to change over time. 
We note that as the node-specific effects $a_i(t),b_j(t)$ go to zero, and sender latent effect $X^s(t)$ and receiver latent effect $X^r(t)$ become identical, our proposed model would degenerate to that of \citet{durante2014bayesian,durante2014nonparametric}.

We specify independent 
Gaussian process priors for $\mu$ and $\beta$:
\begin{gather*}
    \mu(\cdot) \sim GP(0,c_{\mu}),\quad c_{\mu}(t,t')=exp\{-k_{\mu}(t-t')^2\}\\
    \beta_p(\cdot) \sim GP(0,c_p),\quad c_p(t,t')=exp\{-k_p(t-t')^2\}
\end{gather*}

The prior on the latent effects includes temporal dependence and within-unit dependence, but assumes independence across units. That is, for the additive ``sociability'' and ``popularity'' latent effects we assume that $(a_i(t_1),\dots,a_i(t_N),b_i(t_1),\dots,b_i(t_N))$ jointly follows a Gaussian process 
$$\{a_i(t_1),\dots,a_i(t_N),b_i(t_1),\dots,b_i(t_N)\}^T \sim N_{2N}(0,\left( \begin{smallmatrix} c_{a} & \rho_{ab} c_{ab}\\ \rho_{ab} c_{ab}&c_{b} \end{smallmatrix} \right))$$
For the multiplicative latent effects, we also assume Gaussian process but include a shrinkage parameter $\tau$ proposed by \citet{durante2014nonparametric}, allowing for learning the dimension of the latent space. As the dimension of the latent space $h\in \{1,\dots,H^*\}$ increases, $\tau_h^{-1}$ will decay to zero, \begin{equation}
    \tau_h=\prod_{l=1}^h\nu_l,\quad \nu_l \sim Ga(a,1),
    \label{equ:tau}
\end{equation}
limiting the effects of the corresponding latent factors $(x_{\cdot h}^s,x_{\cdot h}^r)$: 
\begin{equation}
    \{x_{ih}^s(t_1),\dots,x_{ih}^s(t_N),x_{ih}^r(t_1),\dots,x_{ih}^r(t_N)\}^T \sim N_{2N}(0,\tau_h^{-1}\left( \begin{smallmatrix} c_{x^s}&\rho_x c_{x^sx^r}\\ \rho_x c_{x^sx^r}&c_{x^r} \end{smallmatrix} \right))
    \label{equ:x}
\end{equation}

For flexibility, we allow for different time-invariant correlation, $\rho_{ab}$,  between ``sociability'' and ``popularity''  and a time-invariant correlation $\rho_x$ between the latent positions of an individual as a sender and a receiver. $\rho_{ab}$ may be relatively large, since an individual willing to make friends may also attract friends while $\rho_x$ may not have such a pattern. With larger correlations $\rho_{ab}$ and $\rho_x$, we expect a more ``symmetric'' socio-matrix due to the similar behavior of individual $i$ as a sender and a receiver. Finally the model naturally captures temporal reciprocity via the Gaussian Process specification, alleviating the need for specifying an additional reciprocity parameter (as is done in \cite{hoff2015dyadic}, for example, which is often difficult to learn).

For simplicity, we assume common characteristics length-scale for additive and multiplicative effects respectively:
\begin{gather*}
  c_{a}(t,t')=c_{b}(t,t')=c_{ab}(t,t')=exp\{-k_{ab}(t-t')^2\}\\
  c_{x^s}(t,t')=c_{x^r}(t,t')=c_{x^sx^r}(t,t')=exp\{-k_{x}(t-t')^2\}
\end{gather*}

Let $\theta$ be the collection of all unknown parameters and latent positions. Sampling from the posterior $p(\theta|Y(1),\dots,Y(t_N))$ proceeds via Gibbs sampling with P\'olya-gamma data augmentation  \citep{polson2013bayesian}. Updating schemes for $\mu,\beta_p,y_{ij}$ and the P\'olya-gamma augmented variable $w_{ij}$ are natural extensions of \cite{durante2014bayesian,durante2014nonparametric}'s sampling scheme. However, the multiplicative and additive latent effects and shrinkage parameter $\nu$ require a new approach. We update them jointly with respect to the unit level due to the within-unit dependence that exists because of the directed nature of the data. See the Appendix for the complete derivation of the sampler. 

\section{Analysis of Official Visitation Networks}
\label{sec:app}
We analyzed the quarterly visiting activities among the 18 largest economies in the world, excluding China and Saudi Arabia, from 2007 to 2016. Specifically, we constructed a dynamic directed network $\{Y_{ij}(t): t \in \{1,\dots,40\},i,j \in \{1,\dots,18\}, i\neq j\}$ based on ``make a visit'' events from Integrated Crisis Early Warning System (ICEWS) Dataverse \citep{DVN/28075_2018}. Here, $Y_{ij}(t)=1$ indicates that country $i$ visited country $j$ at least once during quarter $t$, while $Y_{ij}(t)=0$ indicates that no visit occurred. We model these visits using a variation on the gravity model of \citet{isard1954location} by including the logarithms of the gross domestic product (GDP) of both countries in the previous quarter as predictors of quarter $t$ visits.\footnote{We considered quarterly GDP estimated using the expenditure approach (B1\_GE) and under the CQRSA measurement (National currency, current prices, quarterly levels, seasonally adjusted) from OECD dataset \citep{/content/data/data-00017-en}. We calculated real GDP in dollars using the deflator and exchange rate datasets from the International Monetary Fund (IMF) database \citep{imfgdp}. For Brazil, India, Indonesia, Russia, and Turkey, we used the deflators without seasonable adjustments.} The complete model specification is presented below.
\begin{equation}
    S_{ij}(t)=\mu(t)+\log \text{GDP}_{i,t-1}\beta_1(t)+\log \text{GDP}_{j,t-1}\beta_2(t) + {x^s_i}^T(t)x^r_j(t) + a_i(t) + b_j(t)
\end{equation}

In the presented analysis we set the dimension of the latent space $H^*$ to be 10, the length scales of Gaussian processes $k_{\mu},k_x,k_{\beta_1},k_{ab}$ to be 0.1, the correlation between the latent effects $\rho_x=\rho_{ab}$ to be 0.5, and the shrinkage hyper-parameters $a$ to be 2. To obtain posterior samples, we ran the MCMC sampler for 50,000 iterations, discarding the first 5,000 samples as burn-in, and saving samples every 10 iterations. This resulted in a final sample size of 4,500. The effective sample sizes of $\pi_{ij}(t)$ were mostly around 4,500, indicating good mixing. The posterior mean $\hat{\tau}_h^{-1}$ was 0.3987 for $h=1$, decreasing to 0.2774 for $h=5$, and further to 0.0424 for $h=10$. The AUCs for estimation and prediction were 0.88 and 0.77, respectively, indicating good model performance.

Furthermore, we performed extensive sensitivity analyses to assess the impact of several hyper-parameter choices on the model's performance. The results indicate that the model's performance (both in sample and out of sample) is not significantly influenced by hyper-parameter choice (see Table~\ref{tbl:reals_auc} in the appendix).

\begin{figure}
\includegraphics[width=\linewidth]{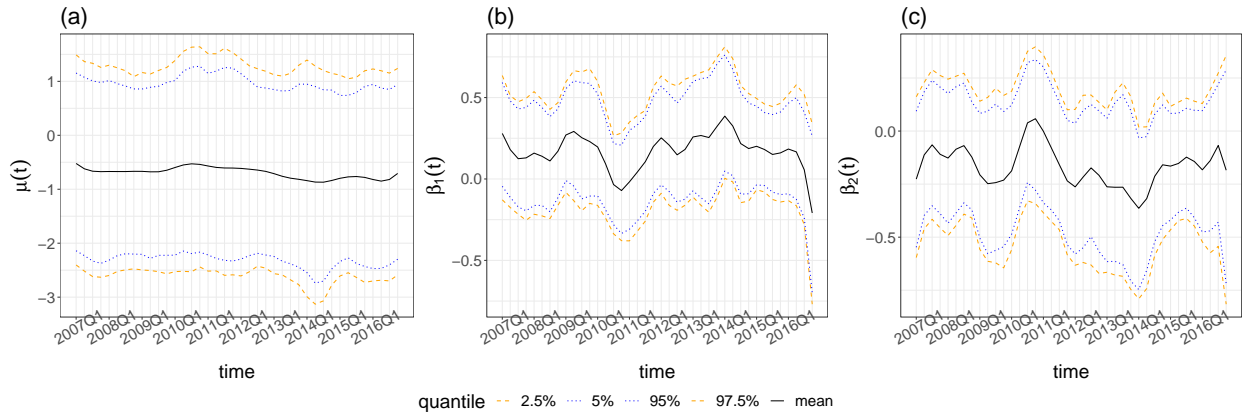}
\caption{Inference based on our model. (a)-(c) indicate graphical summaries of trajectories of posterior samples $\hat{\mu}$, $\hat{\beta}_1$ and $\hat{\beta}_2$. The solid line represents the posterior mean. The grey ribbon represents 95\% highest posterior density interval.}
\label{fig:mubeta12}
\end{figure}

\begin{figure}
\includegraphics[width=\linewidth]{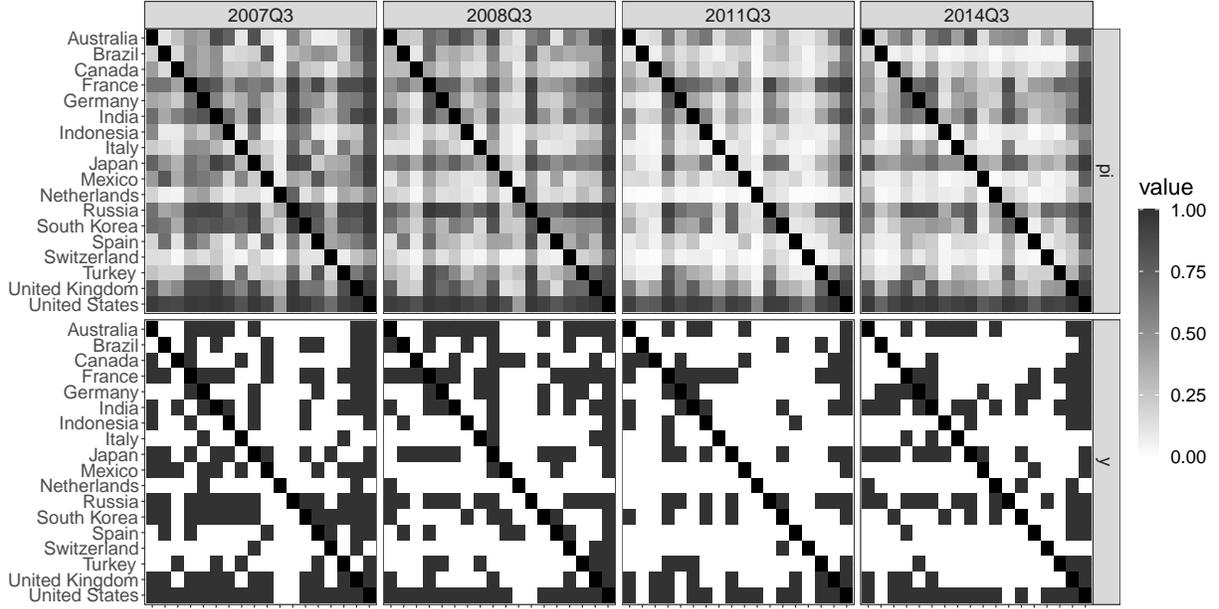}
\caption{Model performance at selected time points. The top panel indicates posterior mean $\hat{\pi}(t)$ and the bottom panel indicates the true relation data. From left to right, each panel indicates a matrix at the different time quarters.}
\label{fig:pimat42}
\end{figure}

\subsection{Discussion of results}
Fig.~\ref{fig:mubeta12} illustrates the baseline trend $\mu(t)$ and the effect of previous quarter's gross domestic product of the sending and receiving country ($\log \text{GDP}_{i,t-1}, \log \text{GDP}_{j,t-1}$). The flat $\mu(t)$ line indicates that visits between the 18 countries remained relatively stable, without any significant trends or sudden changes. The posterior mean estimates of $\hat{\beta}_1(t)$ and $\hat{\beta}_2(t)$ possess opposite signs but similar absolute values, suggesting that countries with a higher GDP tend to prefer making visits over hosting visits, and larger economic disparities between the visiting and hosting countries result in increased probability of connections. This demonstrates that countries with greater resources are generally more popular and attractive to those with fewer resources. In addition, the V-shaped trend of $\hat{\beta}_1(t)$ (or inverse V-shaped of $\hat{\beta}_2(t)$) from 2009Q2 to 2012Q1 implies that the influence of economics on visitation decreased from 2009 to 2010 and then rebounded, likely corresponding to the global financial crisis that began in 2008.

Fig.~\ref{fig:pimat42} displays the relational matrices between 18 countries at selected time points, with darker colors indicating a higher probability of an edge. The greyscale patterns in both the top and bottom panels are very similar, suggesting that our model accurately captures the visiting activity pattern. However, our model fails to capture the pattern for Russia, as Russia frequently visited other countries but was rarely visited by others. This poor fitting performance for Russia may be due to the enforced strong correlation between sociability and popularity of a country ($\rho_{ab}$), which may not be suitable for Russia. 

\begin{figure}[!ht]
\includegraphics[width=\linewidth]{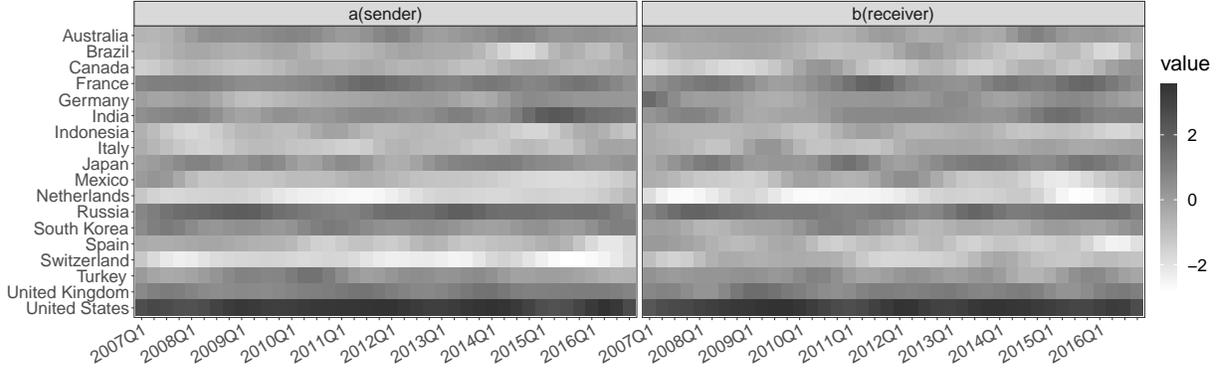}
\caption{Parameter estimation for $a$ and $b$. The left and right panels indicate posterior means for $a$ and $b$ respectively. The rows of each matrix represent different countries while columns represent different time points.}
\label{fig:ab42}
\end{figure}

\begin{figure}[!ht]
\includegraphics[width=\linewidth]{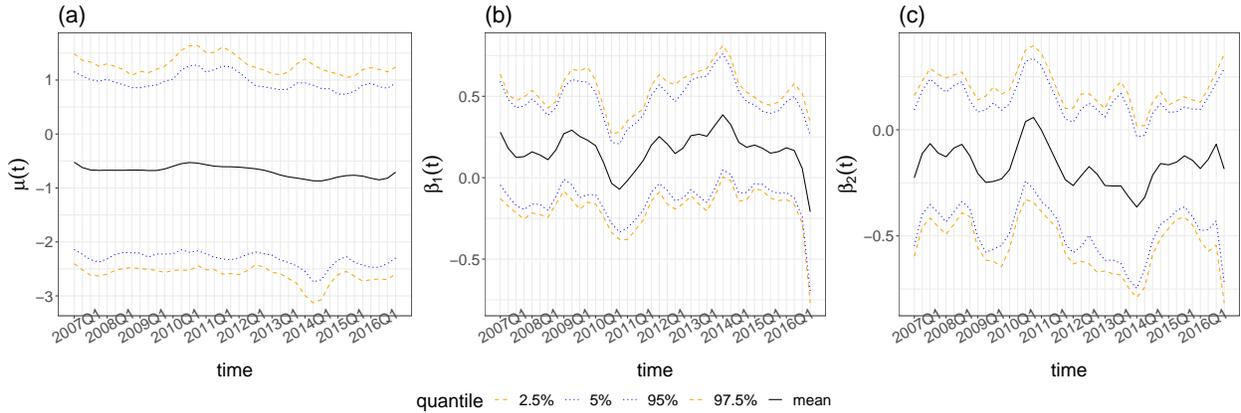}
\caption{Parameter estimation for $a$ and $b$ of Japan. The left and right panels indicate the summary of trajectories for $a$ and $b$ respectively. The solid line represent the posterior mean. The grey ribbon represents 95\% highest posterior density intervals.}
\label{fig:pabt42}
\end{figure}

Fig.~\ref{fig:ab42} summarizes the popularity and sociability parameters, providing a measure of a country's diplomatic role. The United States appears to have played the most active role, followed by Russia, the United Kingdom, France, and Japan, as indicated by high sociability and popularity throughout. In contrast, the Netherlands seems to have played the role of a visitor or host much less frequently than other nations. Fig.~\ref{fig:ab42col} in the Appendix is a colored version of this plot that captures information about the 95\% credible intervals for the elements of Fig.~\ref{fig:ab42}.

Some shifts in the geopolitical roles of certain countries can also be identified. For instance, since Prime Minister Narendra Modi came to power and the economy grew, India has been involved in more diplomatic relationships since 2014Q2. Switzerland became more popular during 2008Q3 compared to other time, which may be attributed to the financial crisis in 2008, as other countries sought Switzerland's financial stability. 
Figure~\ref{fig:lag1} showcases our model's capability to capture temporal reciprocity by displaying the credible intervals for the correlation between $S_{ij}(t_j)$ and $S_{ji}(t_{j-1})$. The pairs $(i,j)$ are ordered according to their posterior mean, and the top 20 country pairs are listed in Table~\ref{tbl:top20}.

\subsection{In depth case study: Japan}

Fig.~\ref{fig:pabt42} illustrates that Japan's sociability, which corresponds to their propensity to visit other countries, dropped following the financial crisis in the second quarter of 2008. The number of countries Japan visited oscillated with various regime changes from 2009 to 2011 (Taro Aso (2008Q3), Yukio Hatoyama(2009Q3), Naoto Kan(2010Q2)). In the second quarter of 2011, Japan reduced its activity significantly, likely due to the Fukushima Daiichi nuclear disaster that occurred in March of the same year. With Prime Minister Abe's return to power and the resulting economic recovery, Japan's sociability has steadily increased since 2012Q2, until a second economic downturn hit in 2014Q2, which was the most substantial decline since the March 2011 earthquake and tsunami. 

The measure of popularity showcases a similar trend.
However, the popularity appears to be more responsive to Japan's economic status, as reflected by larger amplitudes, and less affected by regime alternation, as seen in the relatively stable period from 2009 to 2010. The falls and rises in the popularity curve correspond to recessions (such as financial crisis and disaster) and subsequent recoveries, respectively. 

\begin{figure}[!ht]
\includegraphics[width=\linewidth]{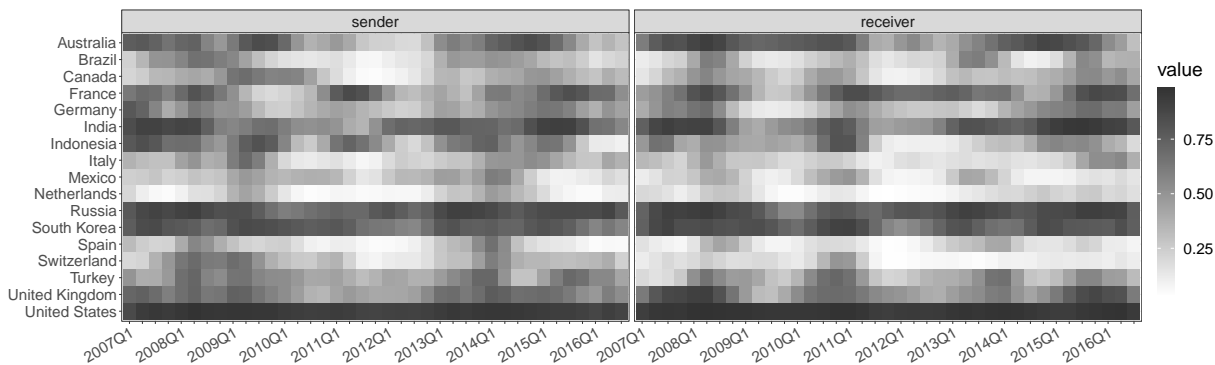}
\caption{Parameter estimation of $\pi_{ij}(t)$ for Japan. The left and right panels indicate posterior means of $\pi_{ij}(t)$ with Japan as a sender (making visits to other countries) and a receiver (hosting visits from other countries) respectively. The rows for each matrix represent different countries while columns represent different time points.}
\label{fig:pijij}
\end{figure}

\begin{figure}[!ht]
\includegraphics[width=\linewidth]{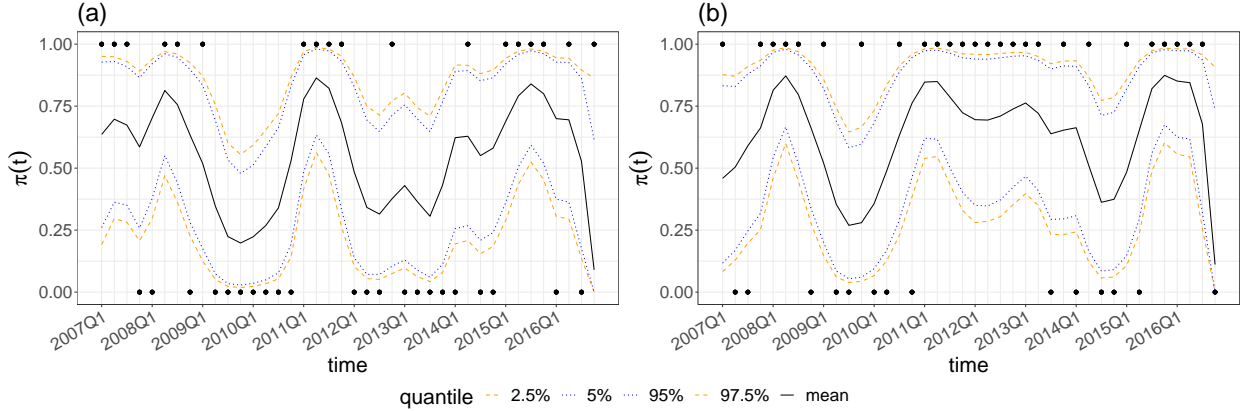}
\caption{Parameter estimation for $\pi_{ij}(t)$ of bilateral relation between Japan and France. The left panel indicates the summary of trajectories for probability of making a visit from Japan to France and vice versa for the right panel. The solid lines represent the posterior mean. The grey ribbons represent 95\% highest posterior density intervals. The black points represent the true relation data.}
\label{fig:pi94}
\end{figure}

To delve deeper into the dynamic bilateral relations, we present the $\pi_{ij}(t)$ matrices for Japan. Fig. \ref{fig:pijij} shows the probability of the existence of bilateral relations, enabling us to identify the countries maintaining a close relationship with Japan and how these relationship have evolved over time. The United States, Russia, and South Korea emerged as Japan's closest partners, while India and Australia appeared as trade partners due to their highly correlated economic trends with Japan. The financial crisis in 2008 and the Fukushima disaster in 2011 led to a systematic decrease in the interactions between Japan and all other countries, while Abe’s return and economic recovery brought an increase in these interactions. It is worth noting that after the Fukushima nuclear disaster, France, a close collaborator of Japan in nuclear energy generation, provided long-term assistance to Japan, as indicated by the dark areas from 2011 to 2014 shown in Fig.~\ref{fig:pijij}(b). A detailed trajectory plot shown in Fig. \ref{fig:pi94} unveils the details of the bilateral relation between Japan and France. Following the Fukushima disaster in March 2011, Japan visited France immediately, and France reciprocated by providing long-term support to Japan, possibly highlighting their willingness to assist Japan's recovery, given their own dependence on nuclear power. Fig.~\ref{fig:pijijcol} in the Appendix provides a colored version of this plot that captures information about the 95\% credible intervals for the elements of Fig.~\ref{fig:pijij}.

A similar analysis can be performed for other countries. 
Generally, the patterns of the parameters are closely tied to various events, such as natural disasters (e.g., floods and earthquakes), financial crises, changes in political leadership, contests, terrorism, and other major events like the Olympics. Our proposed model effectively captures the underlying patterns of bilateral relations, detects the relevant social events, and demonstrates a significant advantages in interpreting the corresponding parameters.

\section{Simulation}
\label{sec:sim}
\subsection{Validation}\label{ssec:validation}

We follow \cite{durante2014nonparametric}'s simulation to conduct a numerical experiment to check whether the model can accurately recover the true data generating process when hyperparameters are mis-specified. We generated data according to the model in Eq.~\eqref{equ:ber}--\eqref{S_equ}, with 15 nodes and 40 time points. We generated observed covariates $\boldsymbol{Z_1}$ from a Gaussian process $GP(0.5,c_{z_1})$, where $c_{z_1}(t,t')=exp\{-k_{z_1}(t-t')^2\}$. The remaining parameters were set as follows:
\begin{align*}
    \text{intercept: }&\boldsymbol{\mu} \sim GP(0,c_{\mu}),\quad c_{\mu}(t,t')=exp\{-k_{\mu}(t-t')^2\}\\
    \text{coefficient: }&\boldsymbol{\beta_1} \sim GP(0,c_{1}),\quad c_{1}(t,t')=exp\{-k_{1}(t-t')^2\}\\
    \text{(sociability, popularity): }&(\boldsymbol{a_i},\boldsymbol{b_i}) \sim N(0,\left( \begin{smallmatrix} c_{ab} & \rho_{ab} c_{ab}\\ \rho_{ab} c_{ab}&c_{ab} \end{smallmatrix} \right)), \quad c_{ab}(t,t')=exp\{-k_{ab}(t,t')^2\}\\
    \text{multiplicative effects: }&(\boldsymbol{x_{ih}^s},\boldsymbol{x_{ih}^r}) \sim N(0,\left( \begin{smallmatrix} c_{x} & \rho_{x} c_{x}\\ \rho_{x} c_{x}&c_{x} \end{smallmatrix} \right)), \quad c_{x}(t,t')=exp\{-k_{x}(t,t')^2\}
\end{align*}

For this data generating process, we set the same length scale $k_{\cdot}=0.01$ for all Gaussian process priors, the same correlation $\rho_{\cdot}=0.5$ for both additive and multiplicative effects, and a low true dimension of the multiplicative effect $H=2$. We excluded the entire matrix $Y(40)$ to evaluate the performance of model prediction. For the priors, we considered a larger length scale $k_{\cdot}=0.05$, a higher dimension of multiplicative latent effects $H^*=10$, and a relatively small shrinkage parameter $a=2$ for the latent space. 

\begin{figure}[!ht]
\includegraphics[width=\linewidth]{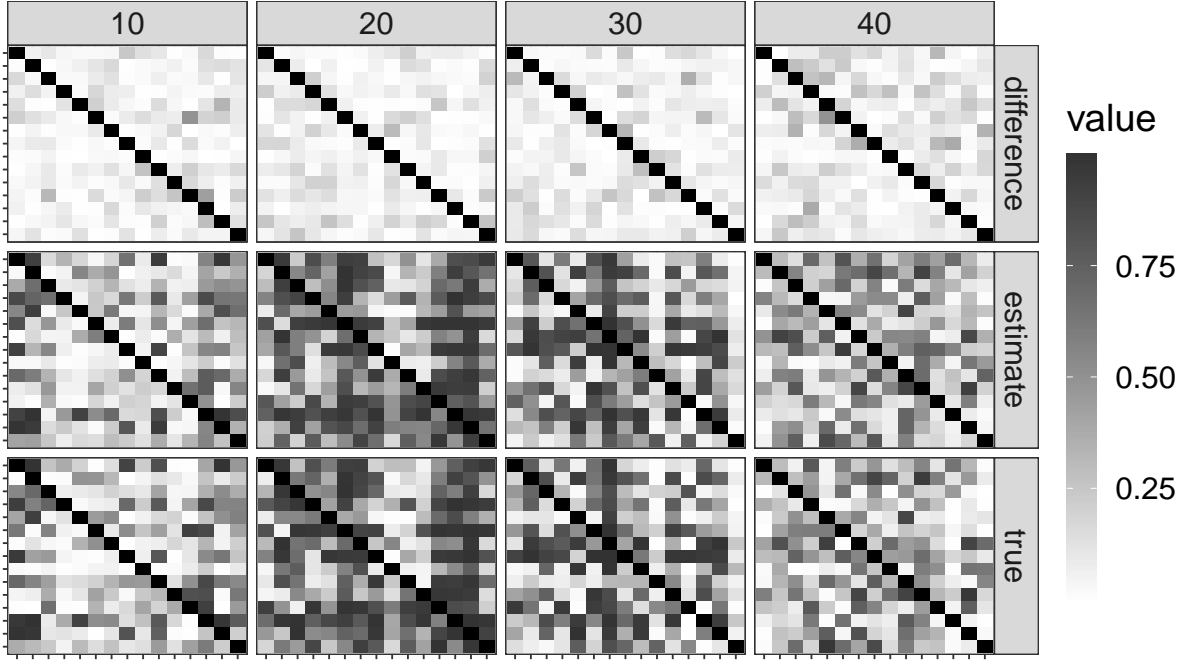}
\caption{Model performance at selected time. From top to bottom, each panel indicates the absolute difference $|\pi(t)-\hat{\pi}(t)|$, estimate values $\hat{\pi}(t)$ and true values $\pi(t)$ respectively. From left to right, each panel indicate matrices at the different time point $t=10,20,30,40$. Dark grey scale indicates high probability of the existence of an edge. Notice the $Y(40)$ is held out for prediction.}
\label{fig:pimat}
\end{figure}

\begin{figure}[!ht]
\includegraphics[width=\linewidth]{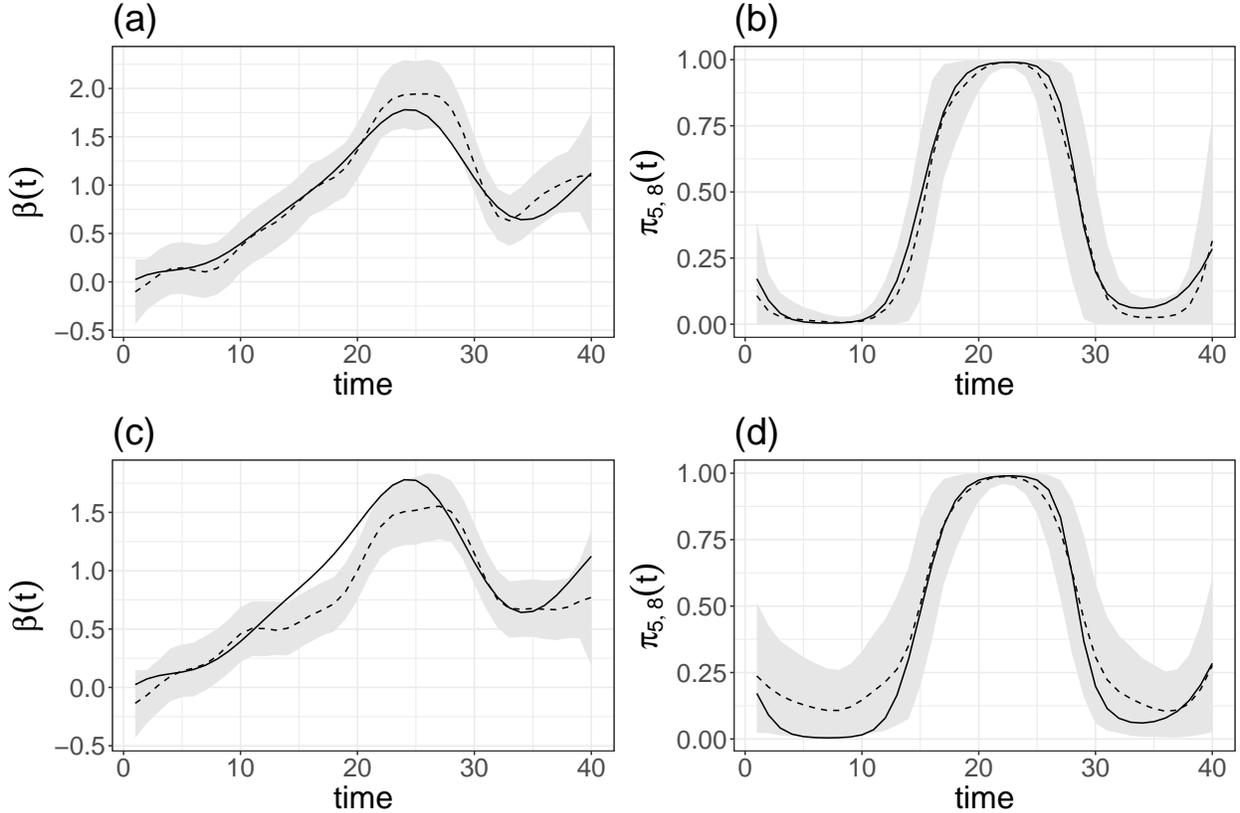}
\caption{Model comparison between the model with network structure and without network structure. (a) and (b) indicate graphical summary of trajectories of sampled $\hat{\beta}$ and $\hat{\pi}_{5,8}$ based on the network structured model, while (c) and (d) indicate those based on the model without network structure. The solid line and dash line represent true and estimated values. The grey ribbon represents 95\% highest posterior density intervals.}
\label{fig:mubetapi}
\end{figure}

We ran 5000 iterations of the MCMC sampler discarding the first 1000 draws as burn-in. The effective sample size of $\pi_{ij}(t)$ is approximately 1500, indicating relatively good mixing. Fig.~\ref{fig:pimat} suggests good performance in model fitting and prediction, with the light color in the first row panels indicating that the estimator recovers the true structure well even for the prediction. The AUCs (Area Under the ROC Curve) for estimation and prediction are 0.9254 and 0.7801, respectively. The shrinkage parameter $\hat{\tau}$ maintains the main effects of the first two latent factors while restricting effects of other factors. As suggested by Eq.~\eqref{equ:x}, smaller $\tau_h^{-1}$ leads to greater shrinkage of the latent factors at the $h$th dimension $(\boldsymbol{x_{\cdot h}^s},\boldsymbol{x_{\cdot h}^r})$ towards $0$. As shown in Table \ref{tbl:tau} in the Appendix, the shrinkage estimator $\hat{\tau}_h^{-1}$ is relatively large for the first two dimensions ($\hat{\tau}_1^{-1},\hat{\tau}_2^{-1} \approx0.91$) and decreases from 0.27 towards 0 for higher dimension $\hat{\tau}_{3:10}^{-1}$. Based on our model, the true parameters lie between 95\% highest posterior density (HPD) intervals and almost overlap with the posterior mean (Fig.~\ref{fig:mubetapi}, Fig.~\ref{fig:ab} and Fig.~\ref{fig:abt}). Furthermore, we compared our model to the model without network structure (simply letting $S_{ij}(t)=\mu(t)+Z_{ij,t}^T\beta(t)$) in terms of the estimation performance of $\hat{\beta}(t)$ and $\hat{\pi}_{5,8}(t)$. Fig.~\ref{fig:mubetapi} shows ignorance of the network structure results in a biased estimator of $\beta(t)$ and a large uncertainty of estimation of $\pi_{5,8}(t)$. Similar patterns are observed for other $\pi_{ij}$.

\subsection{Prior sensitivity}\label{ssec:sensitive}

We conducted simulations with the same settings as before but on larger networks and with more time points. The results for $(N,T)\in\{(30,20),(30,30),(40,20)\}$ showcase that the approach continues to perform well for estimation and prediction (see Table \ref{tbl:vn_auc} in the Appendix). We also conducted extensive sensitivity analysis, considering different values for the dimension of the latent space $H^*$, the correlations between variables $\rho_x,\rho_{ab}$, and the length scales of Gaussian process priors $k_{\mu},k_x,k_{\beta_1},k_{ab}$ (see Table~\ref{tbl:hstar_auc}, Table~\ref{tbl:rho_auc} and Table~\ref{tbl:k_auc} in the Appendix). Generally, a larger $H^*$ brings more flexibility to the model and leads to a better model fitting and prediction (also indicated by Theorem \ref{theorem1} and Corollary \ref{corollary1}). In contrast, underestimating $H$ can cause bias. Therefore, setting a conservative $H^*$ can be a better choice, although it should be balanced against additional computational cost. The simulation results suggest that the model's performance is not overly sensitive to other hyperparameter choices (see Table \ref{tbl:rho_auc} and Table~\ref{tbl:k_auc} in the Appendix). 

\subsection{Model Misspecification}\label{ssec:misspec}

We compare our directed latent space model for networks (DLSN) with the latent space model for directed network (LSMDN) proposed by \cite{sewell2015latent}. Instead of adopting an additive and multiplicative latent factors shown in Eq.~\eqref{S_equ}, \citet{sewell2015latent} introduce radii $r_i$, $r_j$ around an individual to represent ``sociability" and ``popularity", and utilize Euclidean distance to depict the difference between latent factors, where $d_{ijt}=||X_{it}-X_{jt}||$, and $X_{it}$ is the latent factor. Instead of using Gaussian process to model the latent factor, \cite{sewell2015latent} consider a Markov process. Besides the aforementioned two main difference, the LSMDN model does not allow for effects of covariates. 

\begin{align}
    \text{LSMDN: }&S_{ij}(t)=\beta_{IN}(1-\frac{d_{ijt}}{r_j}) + \beta_{OUT}(1-\frac{d_{ijt}}{r_i})\\
    \label{equ:lsmdn}
    \text{DLSN: }&S_{ij}(t)=\mu(t)+Z_{ij,t}^T\beta(t) + {x_i^s}^T(t)x^r_j(t) + a_i(t) + b_j(t)\\
    \text{Naive: }&S_{ij}(t)=\mu(t)+Z_{ij,t}^T\beta(t)\\
    \text{Random Effect: }&S_{ij}(t)=\mu(t)+Z_{ij,t}^T\beta(t) + \epsilon_{ij}(t)
    \label{equ:re}
\end{align}

\begin{figure}
\includegraphics[width=\linewidth]{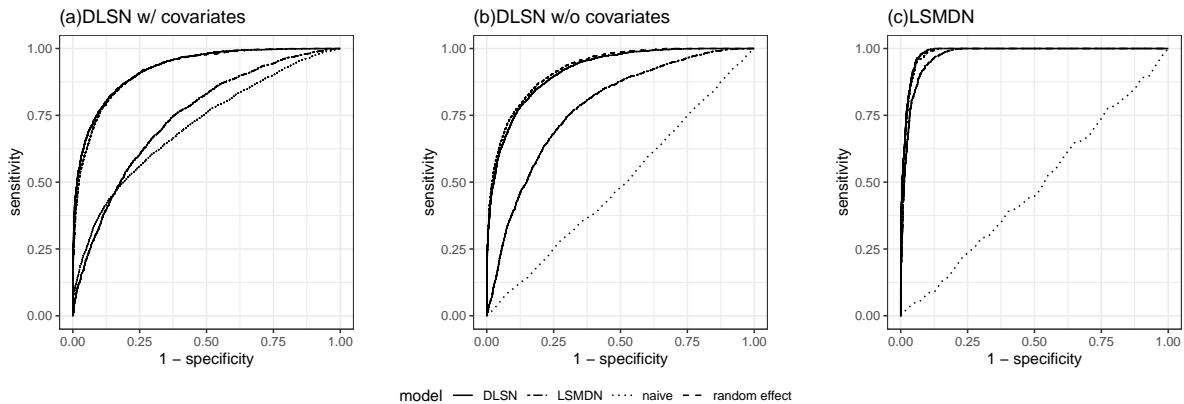}
\caption{ROC curves for model comparison. Different colors represent different data generating processes. Different panels represent different models.}
\label{fig:roccuv}
\end{figure}

We generated three datasets based on the DLSN model with covariates, the DLSN model without covariates, and LSMDN model. We model each of these datasets using four models listed in equations (\ref{equ:lsmdn})-(\ref{equ:re}) : the DLSN model, DLSN model without the network structure (labeled as naive), DLSN model with independent and identically distributed random effects replacing network structure (labeled as random effect) and LSMDN model. The resulting AUCs are shown in Table~\ref{tbl:auc} and ROC curves are shown in Fig.~\ref{fig:roccuv}. Our model exhibits strong performance even when the data are generated from the LSMDN model, and surpasses other models when correctly specifying the true generating process. While our model shares the same level of flexibility as the random effect model, it provides greater interpretability by further decomposing the random effects into meaningful additive and multiplicative latent processes.

\begin{table}
\centering
\begin{tabular}{l|ccccc}
\hline
& DLSN & Random Effect & Naive & LSMDN &\\ 
\hline
DLSN w/ covariates & 0.9257 & 0.9209 &0.7104& 0.7450 &\\
DLSN w/o covariates & 0.9177 & 0.9253 &0.4943& 0.7820&\\ 
LSMDN& 0.9850 & 0.9835 &0.4795& 0.9726 &\\
\hline
\end{tabular}
\caption{Comparison between DLSN and LSMND based on AUCs\label{tbl:auc}. Row names and column names represent data generating process and modeling process respectively.}
\end{table}

\section{Discussion}
\label{sec:discussion}
We extended the nonparametric Bayes dynamic undirected network model of \cite{durante2014nonparametric} and \cite{kim2018dynamic} to the general setting of dynamic additive and multiplicative network (AMEN) models \citep{hoff2015dyadic}. Each effect is modeled using a flexible Gaussian process with a flexible covariance structure, allowing for homogeneity and heterogeneity of sender and receiver effects. As shown in the application of international official visitation networks, our model inherits the advantages in capturing dynamic patterns of dyadic data and shows excellent interpretability. Also, our model is flexible in terms of the capability of modeling both directed and undirected networks (degenerate case) evolving in continuous time and specification of the hyperparameters.

An important direction for future work is accounting for the dynamic evolution of multiple related directed networks. In addition to official visitation dynamic networks, various other types of dynamic networks are observed among the same countries (e.g. expressing intent to cooperate, consulting, engaging in diplomatic or material cooperation, or providing aid). These networks are naturally related, as countries may exhibit similar social behavior across different diplomatic contexts. A natural, albeit simplistic, approach is to assume a shared latent space across different networks, where countries are expected to occupy similar positions in the latent space across networks. Drawing from the concept of Group Factor Analysis \citep{virtanen2012bayesian,klami2014group}, we can decompose the latent factors in Equation~\ref{S_equ} into two parts: common latent factors shared across multiple networks, and distinct group-specific factors for each network. Modeling multiple networks jointly could further facilitate integration of information from multiple networks, enhance our understanding of countries' distinct roles across different relations, and evaluate the impact of worldwide events on social interactions among countries.

The flexibility inherent in the Guassian Process (GP) modeling framework comes at the price of computational complexity. As we employ GPs for both additive and multiplicative latent factors, we are unable to scale to substantially larger networks (though we can scale to larger time-frames).

There is a growing literature on developing scalable Gaussian process methods (see \cite{liu2020gaussian} for a recent review). These methods include estimating the Gaussian process covariance matrix based on a subset of data, using a sparsity matrix or a sparse approximation, and employing the divide-and-conquer approach to obtain local approximation based on partitioned subsets of data. Studying how to implement such improvements for our specific model framework (with the complex dependencies that are necessarily present in network problems) is important future work.

Lastly, we conclude with statements of a theorem and a corollary that demonstrates the suitability and flexibility of our model specification. Theorem \ref{theorem1} and Corollary \ref{corollary1} show that our model specification can capture any true underlying probability matrices of a dynamic directed network.

\begin{theorem}
\label{theorem1}
Given a positive semi-definite matrix $S(t)$, for every $t \in \mathcal{T}$, there exist 

$\{X^s(t),X^r(t), \mu(t),a(t),b(t)\} \in \mathcal{X}_{X^s} \times \mathcal{X}_{X^r}\times\mathcal{X}_{\mu}\times\mathcal{X}_{a}\times\mathcal{X}_{b}$ such that 
$$S(t)=\mu(t)1_V1_V^T+Z^T(t)\beta(t) + X^s(t){X^r}^T(t) + a(t)1_V^T + 1_Vb(t)^T. (t \in \mathcal{T})$$
\end{theorem}

\begin{corollary}
\label{corollary1} Given a asymmetric link probability matrix $\pi(t)$, for every $t \in \mathcal{T}$, here exist 

$\{X^s(t),X^r(t),\mu(t),a(t),b(t)\} \in \mathcal{X}_{X^s}\times\mathcal{X}_{X^r}\times\mathcal{X}_{\mu}\times\mathcal{X}_{a}\times\mathcal{X}_{b}$ such that 
$$\pi_{ij}(t)=\{1+e^{-(\mu(t)+Z_{ij}^T(t)\beta(t) + {x^s_i}^T(t)x^r_j(t) + a_i(t) + b_j(t))}\}^{-1} (i,j=1,\dots,V; i\neq j)$$

\end{corollary}

\section*{Funding}
This research was partially supported by grants from the National Science Foundation (CAREER DMS-2046880); the Army Research Institute (W911NF-18-1-0233).

\section*{Data availability}
Data are available directly from ICEWS \citep{/content/data/data-00017-en}. Code is available on the first author's GitHub.

\clearpage
\newpage
\mbox{~}

\bibliographystyle{chicago}
\bibliography{mybibfile}

\newpage
\appendix
\section*{Supplementary Materials}

\section{Gibbs sampling scheme}
Here we describe the full conditional needed for the Gibbs sampler defined in the main text.
For $i,j=1,\dots,V;i\neq j;h=1,\dots,H^*;p=1,\dots,P$, the algorithm is as follows:

\textbf{Step 1:} Update $w_{ij}(t)$ from its full conditional:
    $$w_{ij}(t)|- \sim PG\{1,\mu(t)+\sum_{h=1}^{H^*}x_{ih}^{sT} x_{jh}^r+Z_{ij,t}^T\beta(t)+a_i(t)+b_j(t)\}$$

\textbf{Step 2:} Update $\mu$ from its full conditional:
    $$\mu|- \sim N(\mu_{\mu},\Sigma_{\mu}),$$
where 
\begin{align*}
    \Sigma_{\mu} &= \{diag(\sum_i\sum_{j\neq i} w_{ij}(t_1),\dots,\sum_i\sum_{j\neq i} w_{ij}(t_N))+K_{\mu}^{-1}\}^{-1}\\
    f^{\mu}_{ij}(t) &= y_{ij}(t)-0.5-w_{ij}(t)(x_i^{sT}(t)x_j^r(t)+Z_{ij}(t)^T\beta(t)+a_i(t)+b_j(t))\\
    \mu_{\mu} &= \Sigma_{\mu}\begin{bmatrix} \sum_i\sum_{j\neq i}f^{\mu}_{ij}(t_1)\\ \vdots \\
\sum_i\sum_{j\neq i}f^{\mu}_{ij}(t_N)
\end{bmatrix}.
\end{align*}

\textbf{Step 3:} Update $\beta$ from its full conditional:
$$\beta_p|- \sim N(\mu_{\beta_p},\Sigma_{\beta_p}),$$
where 
\begin{align*}
    \Sigma_{\beta_p} &= \{diag(\sum_i\sum_{j\neq i} Z_{ij,p}^2(t_1)w_{ij}(t_1),\dots,\sum_i\sum_{j\neq i} Z_{ij,p}^2(t_N)w_{ij}(t_N))+K_p^{-1}\}^{-1}\\
    f^{\beta_p}(t)_{ij} &= Z_{ij,p}(t)(y_{ij}(t)-0.5-w_{ij}(t)(\mu(t_1)+x_i^{sT}(t)x_j^r(t)\\
    &\qquad +Z_{ij(-p)}(t)^T\beta_{(-p)}(t) +a_i(t)+b_j(t)))\\
    \mu_{\beta_p} &= \Sigma_{\beta_p}\begin{bmatrix} \sum_i\sum_{j\neq i}f^{\beta_p}(t_1)_{ij}\\ \vdots \\
\sum_i\sum_{j\neq i}f^{\beta_p}(t_N)_{ij}
\end{bmatrix}
\end{align*}

\textbf{Step 4:} Update $X^{(v)}$ from its full conditional:
Let 
\begin{align*}
    X^{(v)} &=\{x_{v1}^s(t_1),\dots,x_{v1}^s(t_N),\dots,x_{v H^*}^s(t_1),\dots,x_{v H^*}^s(t_N),\\
    &\qquad\qquad x_{v1}^r(t_1),\dots,x_{v1}^r(t_N),\dots,x_{v H^*}^r(t_1),\dots,x_{v H^*}^r(t_N)\}^T,
\end{align*}
then we have 
$$X^{(v)}|- \sim N(\mu_{X^{(v)}},\Sigma_{X^{(v)}}),$$
where 
\begin{align*}
    \Sigma_{X^{(v)}}=\{\Tilde{X}^{(-v)T}\Omega^{(v)}\Tilde{X}^{(-v)}+K_{X^{(v)}}^{-1}\}^{-1}\\
    \mu_{X^{(v)}}=\Sigma_{X^{(v)}}[\Tilde{X}^{(-v)T}\{y^{(v)}-1_{2N(V-1)}0.5-\Omega^{(v)}(1_{2(V-1)}\otimes \mu + (Z\beta)^{(v)}+a^{(v)}+b^{(v)})\}]
\end{align*}
The components of the distribution
are as follows:
\begin{align*}
    \Tilde{X}^{(-v)} &=
\begin{pmatrix}
x_{11}^r(t_1)&x_{12}^r(t_1)&\dots&x_{1H^*}^r(t_1)&&&&\\
\ddots&\ddots&\dots&\ddots&&&&\\
x_{11}^r(t_N)&x_{12}^r(t_N)&\dots&x_{1H^*}^r(t_N)&&&&\\
\vdots&\vdots&\dots&\vdots&&&&\\
x_{V1}^r(t_1)&x_{V2}^r(t_1)&\dots&x_{VH^*}^r(t_1)&&&&\\
\ddots&\ddots&\dots&\ddots&&&&\\
x_{V1}^r(t_N)&x_{V1}^r(t_N)&\dots&x_{VH^*}^r(t_N)&&&&\\
&&&&x_{11}^s(t_1)&x_{12}^s(t_1)&\dots&x_{1H^*}^s(t_1)\\
&&&&\ddots&\ddots&\dots&\ddots\\
&&&&x_{11}^s(t_N)&x_{12}^s(t_N)&\dots&x_{1H^*}^s(t_N)\\
&&&&\vdots&\vdots&\dots&\vdots\\
&&&&x_{V1}^s(t_1)&x_{V2}^s(t_1)&\dots&x_{VH^*}^s(t_1)\\
&&&&\ddots&\ddots&\dots&\ddots\\
&&&&x_{V1}^s(t_N)&x_{V2}^s(t_N)&\dots&x_{VH^*}^s(t_N)
\end{pmatrix}
\end{align*}
captures the sender and receiver multiplicative latent positions for all units other than $v$, 
and
\begin{align*}
    \Omega^{(v)} &= diag(\{w_{v1}(t_1),\dots,w_{v1}(t_N),\dots,w_{vV}(t_1),\dots,w_{vV}(t_N),\\
    &\qquad\qquad  w_{1v}(t_1),\dots,w_{1v}(t_N),\dots,w_{Vv}(t_1),\dots,w_{Vv}(t_N)\}^T),\\
    y^{(v)} &= \{y_{v1}(t_1),\dots,y_{v1}(t_N),\dots,y_{vV}(t_1),\dots,y_{vV}(t_N),\\
     &\qquad\qquad 
    y_{1v}(t_1),\dots,y_{1v}(t_N),\dots,y_{Vv}(t_1),\dots,y_{Vv}(t_N)\}^T, \\
    (Z\beta)^{(v)} & = \{Z_{v1}(t_1),\dots,Z_{v1}(t_N),\dots,Z_{vV}(t_1),\dots,Z_{vV}(t_N),\\
     &\qquad\qquad Z_{1v}(t_1),\dots,Z_{1v}(t_N),\dots,Z_{Vv}(t_1),\dots,Z_{Vv}(t_N)\}^T \times (1_{2(V-1)} \otimes \beta), \\
    a^{(v)} &= \{a_{v}(t_1),\dots,a_{v}(t_N),\dots,a_{v}(t_1),\dots,a_{v}(t_N),\\
     &\qquad\qquad a_{1}(t_1),\dots,a_{1}(t_N),\dots,a_{V}(t_1),\dots,a_{V}(t_N)\}^T, \\
    b^{(v)} &= \{b_{1}(t_1),\dots,b_{1}(t_N),\dots,b_{V}(t_1),\dots,b_{V}(t_N),\\
     &\qquad\qquad b_{v}(t_1),\dots,b_{v}(t_N),\dots,b_{v}(t_1),\dots,b_{v}(t_N)\}^T
\end{align*}
contain the weights, observed outcomes, expected outcomes and additive sender and receiver effects for unit $v$. Lastly,

\begin{align*}
    K_{X^{(v)}}^{-1} &= \{\left( \begin{smallmatrix} 1&\rho\\ \rho &1 \end{smallmatrix} \right) \otimes [diag(\tau_1^{-1},\dots,\tau_{H^*}^{-1})\otimes K_x]\}^{-1}\\
    &=\{\frac{1}{1-\rho ^2}\left( \begin{smallmatrix} 1&-\rho\\ -\rho &1 \end{smallmatrix} \right) \otimes [diag(\tau_1,\dots,\tau_{H^*})\otimes K_x^{-1}]\}
\end{align*}
is a computationally tractable formulation of the inverse of the covariance matrix.

\textbf{Step 5:} Update $\tau$ from its full conditional:
\begin{align*}
    \nu_l|- &\sim Ga(a+N\times V \times (H^*-l+1),1+\frac{1}{2}\sum_{h=l}^{H^*}(\prod_{\begin{smallmatrix} t=1\\ t \neq l \end{smallmatrix}}^h \nu_t)\sum_{i=1}^V X_{il}^{*T}\left( \begin{smallmatrix} \Sigma_x&\rho_x \Sigma_x\\ \rho_x \Sigma_x&\Sigma_x \end{smallmatrix} \right)^{-1}X_{il}^*)\\
    \tau_h &=\prod_{l=1}^h\nu_l \textrm{ where } X^*_{il}=\{x_{il}^s(t_1),\dots,x_{il}^s(t_N),x_{il}^r(t_1),\dots,x_{il}^r(t_N)\}^T
\end{align*}

\textbf{Step 6:} Update $(a^{(v)},\textrm{ and } b^{(v)})$ from their full conditionals:

Writing $(a^{(v)},b^{(v)})=\{a_{v}(t_1),\dots,a_{v}(t_N),b_{v}(t_1),\dots,b_{v}(t_N)\}^T$, 
we have
$$(a^{(v)},b^{(v)})|- \sim N(\mu_{ab^{(v)}},\Sigma_{ab^{(v)}})$$
where
\begin{align*}
    \Sigma_{ab^{(v)}} &= \{diag(\sum_{j \neq v} w_{vj}(t_1),\dots,\sum_{j \neq v} w_{vj}(t_N),\sum_{i \neq v} w_{iv}(t_1),\dots,\sum_{i\neq v} w_{iv}(t_N))+K_{ab^{(v)}}^{-1}\}^{-1}\\
    K_{ab^{(v)}}^{-1} &= \{\left( \begin{smallmatrix} 1&\rho_{ab}\\ \rho_{ab} &1 \end{smallmatrix} \right) \otimes \Sigma_{ab}\}^{-1}=\left( \begin{smallmatrix} 1&\rho_{ab}\\ \rho_{ab} &1 \end{smallmatrix} \right)^{-1} \otimes \Sigma_{ab}^{-1}
\end{align*}
and 
\begin{align*}
    \mu_{ab^{(v)}} &= \Sigma_{ab^{(v)}}\begin{bmatrix} \sum_{j \neq v}f^{\mu_{ab}}_b(t_1)_{vj}\\ \vdots \\
    \sum_{j \neq v}f^{\mu_{ab}}_b(t_N)_{vj}\\
    \sum_{i \neq v}f^{\mu_{ab}}_a(t_1)_{iv}\\ \vdots \\
    \sum_{i \neq v}f^{\mu_{ab}}_a(t_N)_{iv}
    \end{bmatrix}\textrm{ where}\\
    f^{\mu_{ab}}_b(t)_{ij} &= y_{ij}(t)-0.5-w_{ij}(t)[(\mu(t)+x_i^{sT}(t)x_j^r(t)+Z_{ij}(t)^T\beta(t)+b_j(t))\\
    f^{\mu_{ab}}_a(t)_{ij} &= y_{ij}(t)-0.5-w_{ij}(t)[(\mu(t)+x_i^{sT}(t)x_j^r(t)+Z_{ij}(t)^T\beta(t)+a_i(t))
\end{align*}

\textbf{Step 7:} If there are any missing values we can sample them from their full conditionals as well, via:
$$y_{ij}(t)|- \sim Ber[\{1+e^{-(\mu(t)+Z_{ij,t}^T\beta(t)+x^s_i(t)^Tx_j^r(t)+a_i(t)+b_j(t))}\}^{-1}]$$


\section{Proofs of Theorems in Discussion}

\textbf{Proof of Theorem~\ref{theorem1}}

Without loss of generality, we assume that $\mu(t)=\beta(t)=a(t)=b(t)=0$. For any matrix $S(t)$ at any time $t \in \mathcal{T}$ can be expressed as $$S(t)=UDV^T=UD^{1/2}D^{1/2}V^T$$

Denote $X^{s}=UD^{1/2}$, $X^r=VD^{1/2}$. Then we have $S(t)=X^s{X^r}^T$, for any $t \in \mathcal{T}$

\textbf{Proof of Theorem~\ref{corollary1}}

Notice the logistic function is one-to-one continuous increasing function, together with Theorem \ref{theorem1}, the proof is straightforward.

\newpage

\section{Additional figures and tables}
\subsection{Simulations}
We include several additional tables and figures that depict simulation results. Table~\ref{tbl:tau} depicts values for the shrinkage estimator $\hat{\tau}_h^{-1}$ for several choices of $H^*$, demonstrating the flexibility of the model in adapting to different dimensional problems.
\begin{table}[!ht]
\begin{center}
\begin{tabular}{c|cccccccccccc}
\hline
$\tau^{-1}$&1&2&3&4&5&6&7&8&9&10\\
\hline
$H^*=10$&0.915&0.910&0.272&0.08&0.025&0.008&0.002&0.001&0&0\\
$H^*=5$&0.930&0.233&0.182&0.83&0.204\\
$H^*=2$&0.952&0.866\\
\hline
\end{tabular}
\end{center}
\caption{Parameter estimation. Posterior mean of $\tau^{-1}$ based on different $H^*$ (True $H=2$) \label{tbl:tau}}
\end{table}

\begin{figure}[!ht]
\includegraphics[width=\linewidth]{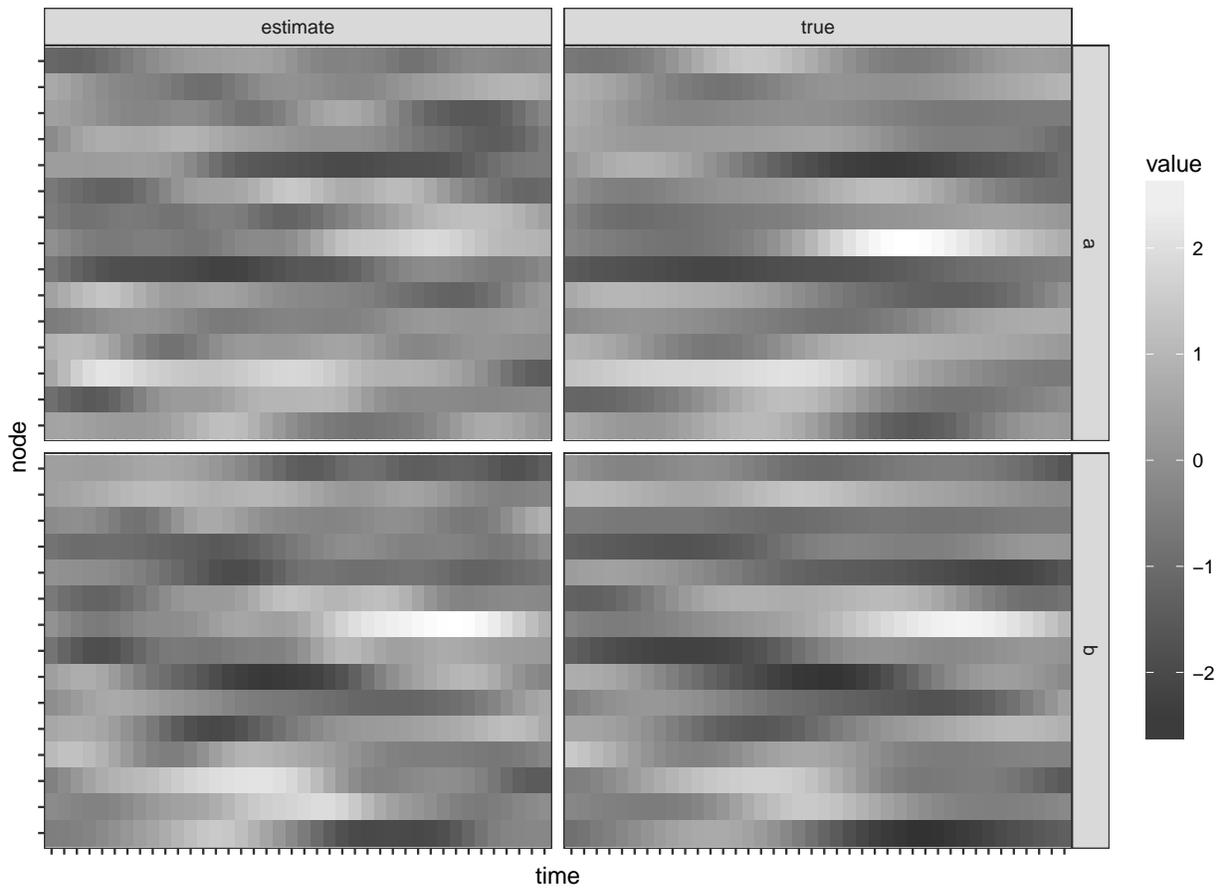}
\caption{Parameter estimation for $a,b$. Top and bottom panels indicate matrices for $a$ and $b$ respectively. Left and right panels indicate estimate and true values respectively. The rows for each matrix represent different nodes while columns represent different time point.}
\label{fig:ab}
\end{figure}

\begin{figure}[!ht]
\includegraphics[width=\linewidth]{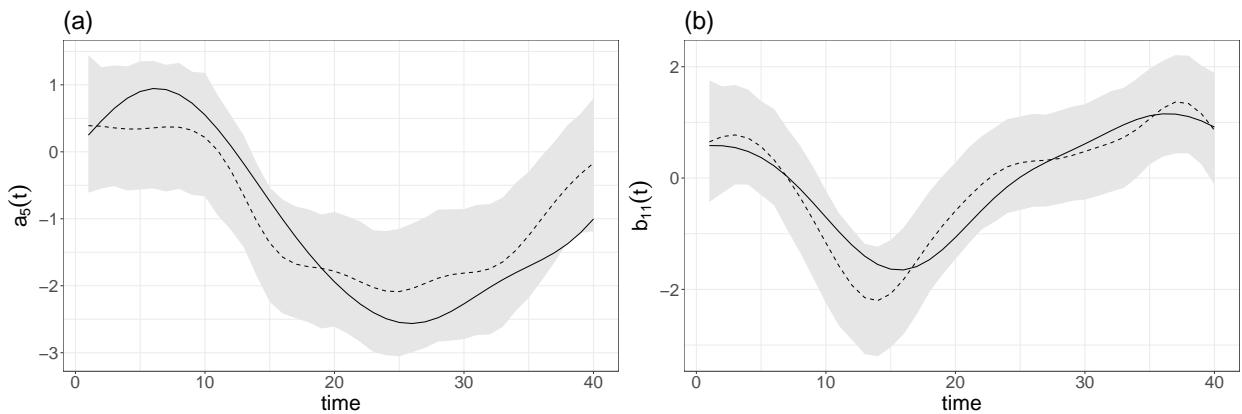}
\caption{Parameter estimation for $a_5(t)$ and $b_{11}(t)$. (a) and (b) indicate graphical summarise of trajectories of sampled $a_5(t)$ and $b_{11}(t)$ respectively. The solid line and dash line represent true and estimated values. The grey ribbon represents 95\% highest posterior density intervals.}
\label{fig:abt}
\end{figure}

Figures~\ref{fig:ab} provides visual validation of the ability of the model to decompose the signal into additive sender and receiver effects. Figure~\ref{fig:abt} showcases the 95\% credible intervals for individual $a$ and $b$.

\subsection{Sensitivity analysis}
This section includes Tables~\ref{tbl:hstar_auc}-\ref{tbl:reals_auc} that describe the sensitivity of the proposed approach to choices of latent dimension (versus true latent dimension), choice of prior correlation parameter, choice of scale of the Gaussian process prior, dimension of the problem and hyperparameter choice in the analysis of the international relations data-set.
\begin{table}[H]
\begin{center}
\begin{tabular}{cc|cc}
\hline
$H$&$H^*$ & auc(fitting)&auc(prediction)\\
\hline
2&2&0.9220&0.7912\\ 
2&5&0.9270&0.6959\\ 
2&10&0.9254&0.7800\\ 
5&2&0.9216&0.8327\\
5&5&0.9532&0.8125\\
5&10&0.9535&0.8658\\
\hline
\end{tabular}
\end{center}
\caption{Sensitivity analysis of the dimention of latent space $H^*$.\label{tbl:hstar_auc}}
\end{table}
\begin{table}[!ht]
\begin{center}
\begin{tabular}{cccc|ccc}
\hline
 $\rho_x$(DGT)&$\rho_{x0}$(fit)&$\rho_{ab}$(DGT)&$\rho_{ab0}$(fit)&auc(fitting)&auc(prediction) \\
\hline
0.5&0.5&0.5&0.5&0.9254&0.7801\\
0.8&0&0.5&0.5&0.9254&0.7801\\
0&0.5&0.5&0.5  &0.9284&0.8242\\
0.5&0.5&0.8&0&0.9264&0.8626\\
0.5&0.5&0&0.5  &0.9262&0.8290\\
0.8&0&0.8&0  &0.9299&0.8033\\
0&0.5&0&0.5    &0.9260&0.8240\\
0.8&0&0&0.5  &0.9291&0.7650\\
0&0.5&0.8&0    &0.9331&0.8676\\
\hline
\end{tabular}
\end{center}
\caption{Sensitivity analysis of the correlation between sender effect and receiver effect $\rho_x$,$\rho_{ab}$ \label{tbl:rho_auc}. DGT represents data generating process.}
\end{table}
\begin{table}[!ht]
\begin{center}
\begin{tabular}{c|ccc}
\hline
$k$ & auc(fitting)&auc(prediction)\\
\hline
0.001&0.8979&0.8761\\
0.01(true)& 0.9117&0.8468\\
0.05&0.9254&0.7801\\
0.1&0.9345&0.8087\\
\hline
\end{tabular}
\end{center}
\caption{Sensitivity analysis of the length scales of Gaussian process priors $k$.\label{tbl:k_auc}}
\end{table}
\begin{table}[!ht]
\begin{center}
\begin{tabular}{cc|cccc}
\hline
$V$ & $N$ & AUC(fitting) & AUC(prediction)\\
\hline
15&40&0.8512&0.9270\\
30&20&0.8027&0.9043\\
40&20&0.8242&0.9112\\
30&30&0.8257&0.8956\\
\hline
\end{tabular}
\end{center}
\caption{Sensitivity analysis of different nodes and time points settings. \label{tbl:vn_auc}}
\end{table}
\begin{table}[!ht]
\begin{center}
\begin{tabular}{ccc|cccc}
\hline
$H^*$ & $a$ &$k$ & AUC(fitting)&AUC(prediction)\\
\hline
5&2&0.1&0.8792&0.7685\\
10&2&0.1&0.8865&0.7688\\
5&10&0.1&0.8680&0.7666\\
5&2&0.01&0.8748&0.8405\\
\hline
\end{tabular}
\end{center}
\caption{Sensitivity analysis of the hyperparameter setting for real data analysis. \label{tbl:reals_auc}}
\end{table}

\subsection{Data analysis}
This section includes additional tables and figures supporting the analysis of the international relations data set in Section~\ref{sec:app}.
\begin{table}[!ht]
\begin{center}
\begin{tabular}{cc|ccc}
sender & receiver&mean&lower (5\%)&upper (95\%)\\
\hline
Russia&Spain&0.570&0.215&0.824\\
Mexico&Spain&0.570&0.187&0.836\\
Spain&Mexico&0.555&0.174&0.835\\
Spain&Russia&0.536&0.162&0.817\\
Brazil&Spain&0.526&0.143&0.807\\
Spain&Turkey&0.513&0.118&0.811\\
Spain&Brazil&0.509&0.141&0.791\\
Japan&Australia&0.495&0.112&0.782\\
Turkey&Spain&0.484&0.082&0.800\\
Switzerland&Spain&0.465&0.013&0.832\\
Germany&India&0.463&0.079&0.751\\
France&Spain&0.456&0.064&0.761\\
Spain&United Kingdom&0.455&0.031&0.769\\
India&Germany&0.452&0.065&0.756\\
Russia&Brazil&0.449&0.054&0.753\\
Turkey&Switzerland&0.440&0.001&0.763\\
Spain&Germany&0.435&0.006&0.775\\
Japan&India&0.427&0.029&0.751\\
Mexico&Brazil&0.422&0.041&0.748\\
Germany&Spain&0.417&0.003&0.752\\
Brazil&Mexico&0.408&0.012&0.734\\
France&India&0.401&0.017&0.722\\
\hline
\end{tabular}
\end{center}
\caption{ Top 20 countries pairs with highest temporal reciprocity, represented by posterior mean and  credible intervals for the correlation between $S_{ij}(t_j)$ and $S_{ji}(t_{j-1})$. \label{tbl:top20}}
\end{table}


\begin{figure}[!ht]
\includegraphics[width=\linewidth]{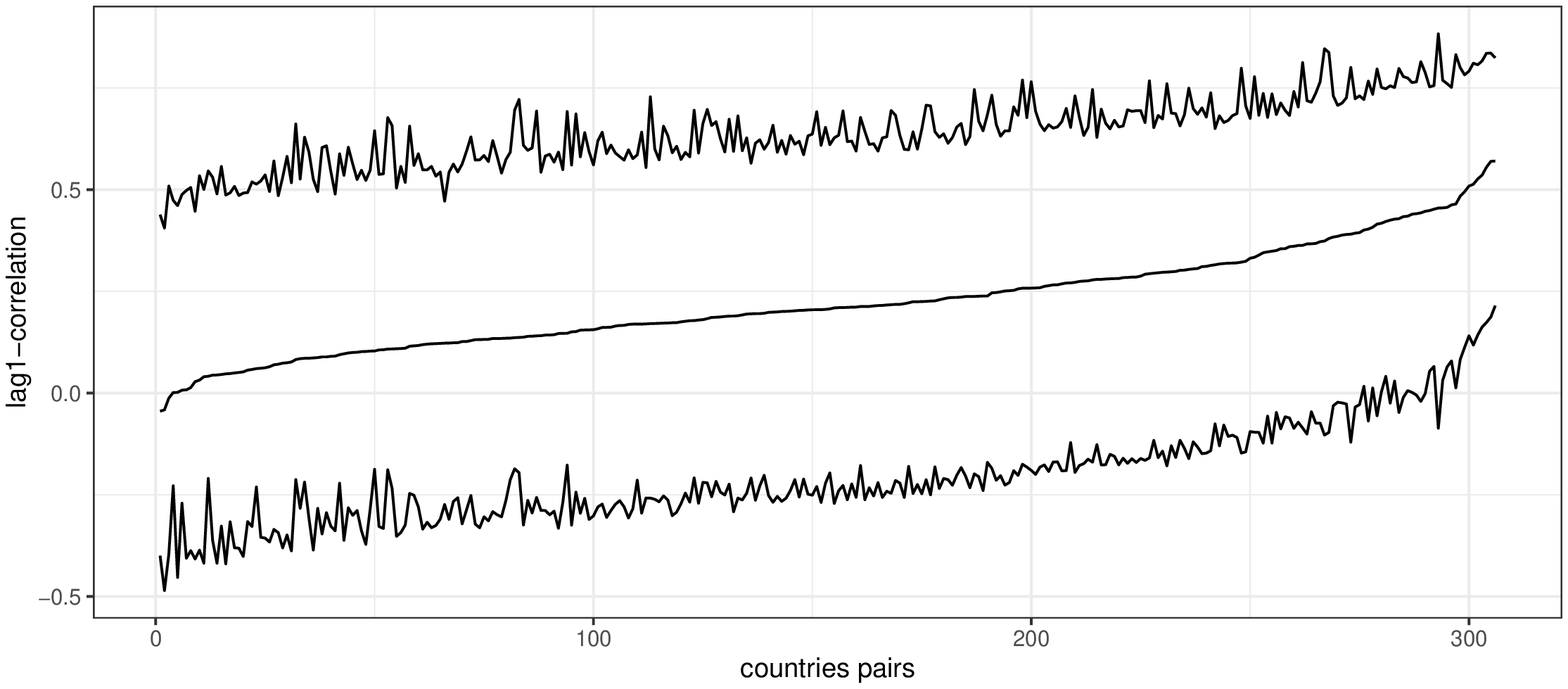}
\caption{Parameter estimation for temporal reciprocity.  we plot the credible intervals for the correlation between $S_{ij}(t_j)$ and $S_{ji}(t_{j-1})$. The pairs $(i,j)$ are ordered by their posterior mean and we list the top 20 country pairs in Table~\ref{tbl:top20}. }
\label{fig:lag1}
\end{figure}

\begin{figure}[!ht]
\includegraphics[width=\linewidth]{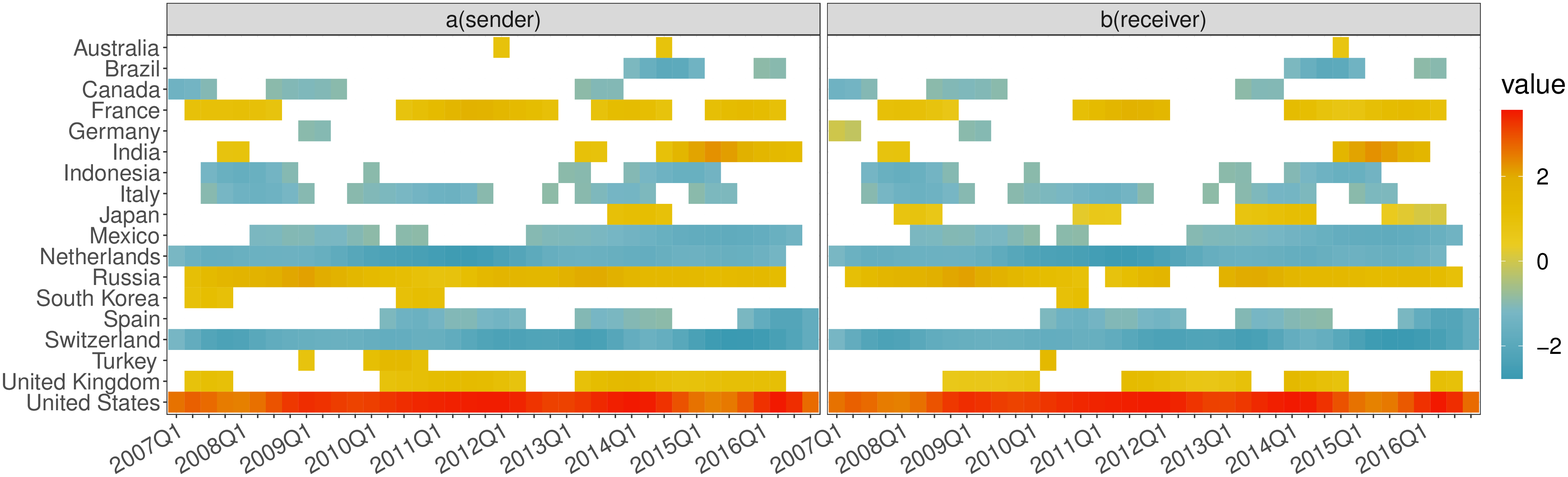}
\caption{Parameter estimation for $a$ and $b$ (plots of values for which the 95\% credible interval does not include 0). The left and right panels indicate posterior means for $a$ and $b$ respectively. The rows of each matrix represent different countries while columns represent different time points.}
\label{fig:ab42col}
\end{figure}

\begin{figure}[!ht]
\includegraphics[width=\linewidth]{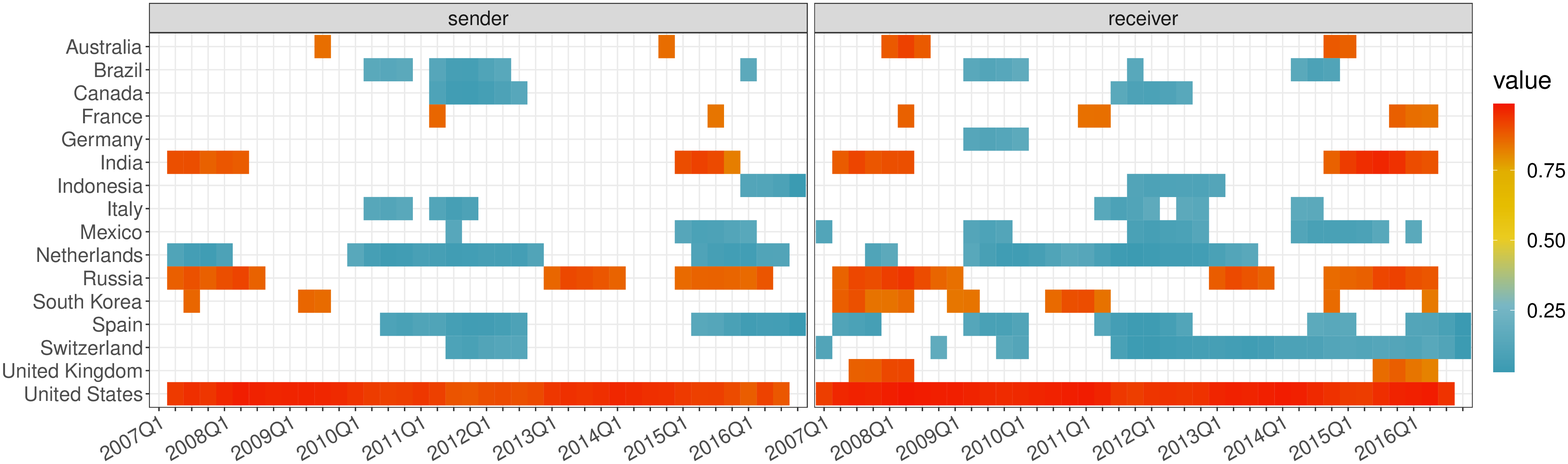}
\caption{Parameter estimation of $\pi_{ij}(t)$ for Japan (plots of values for which the 95\% credible interval does not include 0). The left and right panels indicate posterior means of $\pi_{ij}(t)$ with Japan as a sender (making visits to other countries) and a receiver (hosting visits from other countries) respectively. The rows for each matrix represent different countries while columns represent different time points.}
\label{fig:pijijcol}
\end{figure}

Figures~\ref{fig:ab42col} and \ref{fig:pijijcol} present the estimates of model parameters whenever the 95\% credible intervals for those entries do not include 0. 

\end{document}